\documentclass[usenatbib]{emulateapj}
\usepackage{graphicx}
\usepackage[flushleft]{threeparttable}
\usepackage[usenames,dvipsnames,svgnames,table]{xcolor}
\usepackage{hyperref}
\definecolor{darkblue}{rgb}{0.0,0.0,0.3}
\hypersetup{colorlinks,breaklinks,
            linkcolor=darkblue,urlcolor=darkblue,
            anchorcolor=darkblue,citecolor=darkblue}
\usepackage{amsmath,amssymb}
\usepackage{amsmath}

\begin{document}

\def\etal{et al.\ \rm}
\def\ba{\begin{eqnarray}}
\def\ea{\end{eqnarray}}
\def\etal{et al.\ \rm}
\def\Fdw{F_{\rm dw}}
\def\Tex{T_{\rm ex}}
\def\Fdis{F_{\rm dw,dis}}
\def\Fnu{F_\nu}
\def\FJ{F_J}

\newcommand\cmtrr[1]{{\color{red}[RR: #1]}}
\newcommand\cmtrb[1]{{\color{blue}[RB: #1]}}


\title{$O\left(N^2\right)$ fragmentation algorithm}

\author{Roman R. Rafikov\altaffilmark{1,2,5}, Kedron Silsbee\altaffilmark{3} \& Richard A. Booth\altaffilmark{4}}
\altaffiltext{1}{Department of Applied Mathematics and Theoretical Physics, Centre for Mathematical Sciences, University of Cambridge, Wilberforce Road, Cambridge CB3 0WA, UK}
\altaffiltext{2}{Institute for Advanced Study, 1 Einstein Drive, Princeton NJ 08540}
\altaffiltext{3}{Max-Planck-Institut f\"ur Extraterrestrische Physik, 85748 Garching, Germany}
\altaffiltext{4}{Institute of Astronomy, University of Cambridge, Madingley Road, Cambridge CB3 0HA, UK}
\altaffiltext{5}{E-mail: rrr@damtp.cam.ac.uk}


\begin{abstract}
Collisional fragmentation is a ubiquitous phenomenon arising in a variety of astrophysical systems, from asteroid belts to debris and protoplanetary disks. Numerical studies of fragmentation typically rely on discretizing the size distribution of colliding objects into a large number $N$ of bins in mass space, usually logarithmically spaced. A standard approach for redistributing the debris produced in collisions into the corresponding mass bins results in $\mathcal{O}\left(N^3\right)$ calculation, which leads to significant computational overhead when $N$ is large. Here we formulate a more efficient explicit $\mathcal{O}\left(N^2\right)$ fragmentation algorithm, which works when the size spectrum of fragments produced in an individual collision has a self-similar shape with only a single characteristic mass scale (which can have arbitrary dependence on the energy and masses of colliding objects). Fragment size spectra used in existing fragmentation codes typically possess this property. We also show that our $\mathcal{O}\left(N^2\right)$ approach can be easily extended to work with non-self-similar fragment size distributions, for which we provide a worked example. This algorithm offers a substantial speedup of fragmentation calculations for large $N\gtrsim 10^2$, even over the implicit methods, making it an attractive tool for studying collisionally evolving systems. 
\end{abstract}




\section{Introduction}  
\label{sect:intro}


The issue of collisional fragmentation regularly arises in astrophysical problems where the masses of colliding objects --- e.g. planetesimals or dust particles in protoplanetary disks --- need to be followed. Examples include evolution of the asteroid belt \citep{Durda1997}, the Kuiper Belt \citep{Davis1997,Kenyon2004}, dust populations in debris disks \citep{Kenyon2005,Krivov2008}, and protoplanetary disks \citep{Brauer2008,Birnstiel2010}. The large number of objects in these applications makes it convenient to characterize the state of the system via the mass distribution (spectrum) $n(m)$, such that the number of objects in the mass interval $(m,m+dm)$ is $n(m)dm$. Pair-wise collisions cause mass to be exchanged between different parts of the mass space: a collision between objects $m_1$ and $m_2$ produces a number of fragments, channeling mass towards smaller objects. The total mass in the system of colliding objects is usually conserved in the process, although it may be lost in very energetic collisions when vaporization occurs; for simplicity we will disregard the latter possibility. Also, when particles reach very small sizes they could be removed from the system by other processes such as the Poynting-Robertson drag or radiation pressure. 

In a continuous limit the evolution of the mass spectrum due to fragmentation in pair-wise collisions is described by the following equation:
\ba   
\frac{\partial n(m)}{\partial t} &=& \frac{1}{2}\int dm_1 dm_2 ~g(m|m_1,m_2)
\nonumber\\
& \times & R(m_1,m_2)n(m_1)n(m_2)
\nonumber\\
&- &n(m)\int  dm_1 R(m,m_1)n(m_1).
\label{eq:master_m}
\ea 
Here $R(m_1,m_2)$ is the rate coefficient for collisions between particles of masses $m_1$ (target) and $m_2$ (projectile), while $g(m|m_1,m_2)$ is the size (or mass) distribution of fragments produced in a single collision. It is defined such that the number of fragments in the mass interval $(m,m+dm)$ resulting in a collision between particles $m_1$ and $m_2$ is $g(m|m_1,m_2)dm$. 
In numerical applications the mass coordinate is discretized into a large number $N$ of bins (typically uniformly spaced in $\ln m$). The number of objects per $i$-th bin is $n_i(t)$ and the vector $\vec{n}=\{n_i\}$, $i=1,...,N$ fully characterizes the system. Equation (\ref{eq:master_m}) is then evolved in two steps. First, one chooses a pair of bins $i$ and $j$, $i,j=1,...,N$ and computes the number of collisions between particles in these bins that occur in time $\Delta t$. Second, the fragments produced in collisions of each mass pair are distributed over the $N$ bins according to their mass distribution (i.e. the function $g(m|m_1,m_2)$), which is described by the first term on the right hand side of this equation. This procedure needs to be repeated for each $i,j$ pair of colliding bins.

The first step requires $\mathcal{O}\left(N^2\right)$ operations in general, while the second takes $\mathcal{O}\left(N\right)$. As a result, the numerical cost of evolving equation (\ref{eq:master_m}) scales as $\mathcal{O}\left(N^3\right)$ per time step. This can be rather challenging when $N$ is very large, which is often needed to provide accurate description of collisional evolution of astrophysical systems spanning many orders of magnitude in mass.

The goal of this work is to demonstrate that the numerical cost can be reduced to $\mathcal{O}\left(N^2\right)$ for a certain class of fragment size distributions, which is rather common, and flexible enough to handle even more general models of collision outcomes. We describe this fragmentation model in \S \ref{sect:frag} and the associated $\mathcal{O}\left(N^2\right)$ algorithm in \S \ref{sect:N2frag}. We then show how this model can be extended to approximate more general forms of the fragment size distribution (\S \ref{sect:generalize}) and provide a numerical illustration in \S \ref{sect:piecewise_example}. We compare explicit and implicit methods for evolving fragmentation cascades in \S \ref{sect:implicit}.  Our results are discussed in \S \ref{sect:sum}.


\section{Fragmentation model}  
\label{sect:frag}


The outcome of a collision between two objects depends on a variety of factors. The primary ones are the masses $m_1$ and $m_2$ involved in a collision, the relative velocity of the colliding objects $v_{\rm rel}$, and the material properties of each object determined by the composition, structural characteristics (e.g. porosity) and size. The precise geometry of the collision (i.e. impact parameter for spherical objects) also plays an important role. In this study, to simplify the notation, we will keep track of the dependence of the collision outcome only on the masses of the colliding objects (in many studies the rates of the collisions and collision outcomes are treated in an averaged sense, by convolving over the distributions of $v_{\rm rel}$, impact parameters, etc.).

We characterize the size distribution of fragments forming in a collision of objects with mass $m_1$ and $m_2$ using a reasonably general model of a collision outcome. It covers two most common possibilities, namely (1) the erosion in weakly energetic collisions, which results in one dominant post-collision remnant with the mass $m_{\rm rm}(m_1,m_2)$ and a continuous spectrum of small fragments, and (2) catastrophic disruption, when the large remnant no longer exists and only a continuous spectrum of fragments remains. This model has a form
\ba  
g(m|m_1,m_2) = \epsilon\delta(m-m_{\rm rm}) + g_f(m|m_1,m_2),
\label{eq:coll_outcome}
\ea 
where $\epsilon=1$ in the case of erosion, while $\epsilon=0$ in the case of catastrophic collisions. Here $g_f(m|m_1,m_2)$ is a mass spectrum describing a continuous population of fragments formed in a collision.
There are other possible outcomes of particle collisions --- sticking, mass transfer, bouncing, etc. \citep{Guttler2010,Windmark2012} --- which we do not consider in this study. 

A particular form of $g_f$ explored in this work that allows one to reduce the computational cost of the fragmentation calculation to $\mathcal{O}(N^2)$ is the {\it self-similar} fragment mass spectrum 
\ba
g_f(m|m_1,m_2)=A\varphi\left(\frac{m}{m_*(m_1,m_2)}\right).
\label{eq:ss_model}
\ea
Here $\varphi$ is an arbitrary function that truncates at large masses $m$, $m_*$ is the characteristic mass scale set by $m_1$, $m_2$ and the details of collision physics (i.e. collision energy), and $A$ is the normalization of the spectrum. As we will show later in \S \ref{sect:N2frag}, the key feature of this fragment mass spectrum is that all information about the collision details is absorbed in a single parameter --- the mass scale $m_*$.

The value of $A$ is set by mass conservation (in the absence of mass losses to vaporization)
\ba   
m_1+m_2=\epsilon m_{\rm rm}+A\int_0^\infty m\varphi(m/m_*)dm,
\label{eq:mass_cons}
\ea
so that 
\ba
A=\frac{m_1+m_2-\epsilon m_{\rm rm}}{Im_*^2},~~~~~I=\int_0^\infty z\varphi(z)dz.
\label{eq:A}
\ea
Integration over the mass coordinate can be to extended to infinity since the function $\varphi(z)$ vanishes for large values of $z$. 

Laboratory experiments suggest \citep{Gault1969,Hartmann1969,Fujiwara1977,Blum1993} that the fragment mass spectrum can often be described reasonably well by a power law in fragment mass $m$ truncated above some largest fragment mass $m_*=m_{\rm lf}$:
\ba  
g_f(m|m_1,m_2) &=& A\left\{
\begin{array}{ll}
(m/m_{\rm lf})^\gamma, & ~~~m\le m_{\rm lf}, \\
0,  & ~~~m>m_{\rm lf}. 
\end{array}
\right.
\label{eq:coll_outcome_pl}
\ea 
For example, \citet{Fujiwara1977} found that the mass spectrum of fine fragments resulting in collisions of basaltic bodies can be well described by a power law dependence with index $\gamma=-1.8$. On the other hand, \citet{Blum1993} found that $\gamma=-9/8$ in their experiments with ZrCO$_4$ aggregates. The  spectrum (\ref{eq:coll_outcome_pl}) has the self-similar form (\ref{eq:ss_model}) with 
\ba
\varphi(z)=\varphi^{\rm pl}(z|\gamma)=\Theta(1-z)z^{\gamma},
\ea 
where $z=m/m_{\rm lf}$ and $\Theta(z)$ is the Heavyside step function.


\section{$O\left(N^2\right)$ fragmentation algorithm}  
\label{sect:N2frag}


We now demonstrate how the fragmentation calculation described by the equation (\ref{eq:master_m}) can be turned into an $\mathcal{O}\left(N^2\right)$ problem, rather than $\mathcal{O}\left(N^3\right)$, for the fragment mass spectrum in the form (\ref{eq:ss_model}). We will later show in \S \ref{sect:generalize} that this procedure can be generalized to cover even more complicated fragment mass spectra. 

The basic idea behind this algorithm lies in the order in which different steps are performed. In the standard $\mathcal{O}\left(N^3\right)$ approach for every pair of mass bins the calculation of the collisional debris production is immediately followed by the redistribution step --- assigning fragments to their corresponding mass bin. In our new method the order is different: after computing debris production for each mass bin pair we bin the outcomes (spectrum amplitudes $A$) according to their $m_*$ and $m_{\rm rm}$, possible when $g_f(m|m_1,m_2)$ has a self-similar shape (\ref{eq:ss_model}). Only after this procedure has been carried out for all pairs of bins we perform the fragment distribution step. Both these steps, performed sequentially, can be computed in $\mathcal{O}\left(N^2\right)$ operations, providing the desired speedup. We next describe this algorithm in details. 

Let us introduce two auxiliary $N$-dimensional vectors. One is $\vec{n}^{\rm rm}=\{n^{\rm rm}_i\}$ and is used to record the number of large remnant bodies resulting from erosive collisions in a fixed time interval that end up in the $i$-th bin (i.e. with $m_{\rm rm}$ falling into this bin). Another vector is $\vec{\mathcal{A}}=\{\mathcal{A}_i\}$ --- the sum of normalization factors $A$ given by equation (\ref{eq:A}) for all collisions that have characteristic mass scale of their fragment size distributions $m_*$ falling into the $i$-th mass bin. 

We can now describe our $\mathcal{O}\left(N^2\right)$ fragmentation algorithm step by step. 

\begin{enumerate}

\item At the start of a new time step we set $n^{\rm rm}_i=0$, $\mathcal{A}_i=0$, $i=1,...,N$. 

\item We pick a particular mass bin $i=1,...,N$, which is a $\mathcal{O}(N)$ operation. 

\item We first take care of the last term in the right hand side of equation (\ref{eq:master_m}); although, the order is not important. We consider collisions of objects in the $i$-th bin with objects in every $j=1,...,N$ bins in the system, calculating their rate $R(m_i,m_j)$ and the actual number of collisions in time $dt$:
\ba   
dN^{\rm coll}_{ij}=R(m_i,m_j)n(m_i)n(m_j)dm_i dm_j dt,
\label{eq:dNcoll}
\ea  
where $dm_i$, $dm_j$ are the width of the $i$-th and $j$-th mass bin, respectively. Note that $dN^{\rm coll}_{ij}$ can be non-integer. 

The total loss of particles from the $i$-th bin is then 
\ba
dn^-_i=\sum_{j=1}^N dN^{\rm coll}_{ij}.
\label{eq:dn-}
\ea
Calculation of $dN^{\rm coll}_{ij}$ and $dn^-_i$ for all combinations of $i$ and $j$ requires $\mathcal{O}(N^2)$ operations.

\item We then deal with the first term in the right hand side of equation (\ref{eq:master_m}) and consider the spectrum of fragments resulting in collisions between objects in $i$-th and $j$-th bins considered before. Standard fragmentation algorithms directly distribute these fragments for each $i,j$ pair into the relevant mass bins already at this step (another $\mathcal{O}(N)$ operation). This would make the algorithm scale as $O\left(N^3\right)$. 

We proceed differently. Knowing the relative energy of collision we compute the  values of $m_{\rm rm}$, $m_*$ and $A$ for every pair of $i$ and $j$. We then treat large remnants and small debris as follows.

{\bf Large remnant bodies}\\
We find the index $k_{\rm rm}$ of the bin into which $m_{\rm rm}$ falls, and increase the value of $n^{\rm rm}_{k_{\rm rm}}$ by the number of remnant bodies produced in $dN^{\rm coll}_{ij}$ collisions:
\ba
n^{\rm rm}_{k_{\rm rm}} \to n^{\rm rm}_{k_{\rm rm}} + \epsilon dN^{\rm coll}_{ij}\frac{m_{\rm rm}}{m_{k_{\rm rm}}}.
\label{eq:nrm}
\ea
The factor $m_{k_{\rm rm}}/m_{\rm rm}$ is introduced here to conserve mass: it accounts for the fact that the large remnant mass $m_{\rm rm}$ does not necessarily equal the central mass of the bin $m_{k_{\rm rm}}$. 

{\bf Small fragments}\\
We then take care of the continuous spectrum of smaller fragments. First, we determine the index $k_*$ of the bin into which $m_*$ falls. Since, again, in general $m_{k_*}\neq m_*$, we ensure mass conservation by adjusting $A$ to a (slightly different) value $A^\prime$, such that 
\ba
A^\prime=A\left(\frac{m_*}{m_{k_*}}\right)^2.
\label{eq:Aprime}
\ea
This follows from the fact that the total mass of the self-similar fragment size spectrum with mass scale $m_*$ is $Am_*^2I$, where $I$ is the integral defined in equation (\ref{eq:A}). 

We then update the $k_*$-th component of the vector $\mathcal{A}$ as follows:
\ba
\mathcal{A}_{k_*} \to \mathcal{A}_{k_*}  + dN^{\rm coll}_{ij}A^\prime.
\label{eq:Af}
\ea

\item Operations in steps (2)-(4) are repeated for all pairs of $i$ and $j$ (avoiding double counting). This, in general, requires $\mathcal{O}(N^2)$ calculations, in the end of which vectors $n^{\rm rm}_i$ and $\mathcal{A}_i$ get fully updated. 

\item Now we go through the final, redistribution, steps. We first update the number of objects $n_i$ in each $i=1,...,N$ bins as follows:
\ba
n_i\to n_i+n^{\rm rm}_i+\sum_{j=1}^N \mathcal{A}_j\varphi\left(\frac{m_i}{m_j}\right)dm_i,
\label{eq:update1}
\ea
where $dm_i$ is the width of $i$-th mass bin. In other words, we add to each bin all the large remnants and small fragments that originally fell within its corresponding mass interval. This step again uses $\mathcal{O}\left(N^2\right)$ operations.

Finally, contributions (sinks) from step (3) are subtracted for all $i=1,...,N$ bins:
\ba
n_i\to n_i-dn^-_i,
\label{eq:update2}
\ea
adding $\mathcal{O}\left(N\right)$ additional operations.

\item Time is incremented by $dt$ and steps (1)-(6) are repeated once again. 

\end{enumerate}

One can see that this algorithm indeed performs the fragmentation calculation using only $\mathcal{O}\left(N^2\right)$ operations per time step, and not $\mathcal{O}\left(N^3\right)$ as the conventional approach. This improvement can be achieved only when the collision outcome is described by the equation (\ref{eq:ss_model}), with the mass spectrum of fragments being a self-similar function $\varphi$ with a {\it single characteristic mass scale} $m_*$. 

Indeed, if $\varphi$ depended on e.g. two mass scales, then the amplitude vector $\vec{\mathcal{A}}$ would need to be replaced with the 2-dimensional $N\times N$ amplitude array. In that case the first part of the redistribution step (6) would have involved $\mathcal{O}\left(N^3\right)$ operations, since the summation in equation (\ref{eq:update1}) would need to be carried out over two indices. Similarly, methods based on implicit time-integration are $\mathcal{O}\left(N^3\right)$, see  Section \ref{sect:implicit}. Nevertheless, in the following sections we will show how $\mathcal{O}\left(N^2\right)$ algorithm can be applied also to some more general collision outcomes than the one given by the equation (\ref{eq:ss_model}).


\section{Generalizations of the algorithm}  
\label{sect:generalize}


The method presented in the previous section allows some straightforward generalizations that greatly extend its applicability. Such generalizations are possible when the spectrum of fragments can be represented or approximated using the self-similar components. We cover both cases below.


\subsection{Superposition of self-similar mass spectra}  
\label{sect:many}


A rather straightforward extension of the algorithm outlined above is possible when the fragment mass spectrum can be represented as a sum of $L>1$ self-similar mass distributions:
\ba
g_f(m|m_1,m_2)=\sum\limits_{\eta=1}^L A_\eta\varphi_\eta\left(\frac{m}{m_\eta(m_1,m_2)}\right),
\label{eq:ss_model_sum}
\ea
where characteristic mass scales $m_\eta$ are distinct (i.e. not multiples of each other). In this case amplitudes $A_\eta$ can no longer be found from equation (\ref{eq:A}). Instead, they need to be specified independently, with the only constraint coming from the mass conservation:
\ba   
&& m_1+m_2=\epsilon m_{\rm rm}+\sum\limits_{\eta=1}^L A_\eta m_\eta^2 I_\eta, \\
&& I_\eta=\int_0^\infty z\varphi_\eta(z)dz.
\label{eq:mass_cons_many}
\ea

For example, the full mass spectrum (\ref{eq:coll_outcome}) can be viewed as a sum of two self-similar components: remnant spectrum $\epsilon\delta(m-m_{\rm rm})$ with the amplitude $\epsilon$ and mass scale $m_{\rm rm}$, and the continuous self-similar spectrum of small fragments given by the equation (\ref{eq:ss_model}).

The only difference with the procedure described in \S \ref{sect:N2frag} is that instead of one amplitude vector $\mathcal{A}_i$ we would introduce now $L$ such vectors and then repeat the steps (4)-(6) for all $L$ individual self-similar contributions. The number of operations would scale as $\mathcal{O}\left(LN^2\right)$.


\subsection{Piecewise approximation of the fragment spectrum}  
\label{sect:piecewise}


Our algorithm can also be used when the spectrum of the fragments can be approximated in a piecewise fashion using a number of self-similar components. For example, almost any fragment spectrum can be represented as a series of $S>1$ power law segments (within certain mass intervals) of the form
\ba  
&& g_f(m|m_1,m_2)  \approx  \sum \limits_{s=1}^S \Psi_s(m), \nonumber\\
&& \Psi_s(m) = A_s\left\{
\begin{array}{ll}
0,  & ~m<m^{\rm min}_s,\\
\left(\frac{m^{\rm max}_s}{m}\right)^{\gamma_s}, & ~m^{\rm min}_s\le m\le m^{\rm max}_s, \\
0,  & ~m>m^{\rm max}_s,
\end{array}
\right.
\label{eq:coll_outcome_pw}
\ea 
where $\gamma_s$ is some average of $d\ln g_f(m)/d\ln m$ within the interval $m^{\rm min}_s\le m\le m^{\rm max}_s$. Fragment size distributions in the form of two (or more) broken power laws have been found in collisional experiments of \citet{Fujiwara1977}, \citet{Takagi1984}, \citet{Davis1990}. But any reasonably smooth fragment mass spectrum can be approximated in this way given a sufficiently large number of components (mass intervals) $S$. 

Each of these components can be written as the difference of the two power law spectra $\varphi^{\rm pl}$ defined by equation (\ref{eq:coll_outcome_pl}), namely
\ba
\Psi_s(m) &=& A_s\varphi^{\rm pl}\left(\frac{m}{m^{\rm max}_s}\Big|\gamma_s\right) \nonumber\\
&-& A_s\left(\frac{m^{\rm min}_s}{m^{\rm max}_s}\right)^{\gamma}\varphi^{\rm pl}\left(\frac{m}{m^{\rm min}_s}\Big|\gamma_s\right).
\label{eq:dif}
\ea

Combining equations (\ref{eq:coll_outcome_pw}) and (\ref{eq:dif}) we see that $g_f(m|m_1,m_2)$ ends up being approximated as a linear combination of $2S$ self-similar (power law) components, which reduces the problem to the one already considered in \S \ref{sect:many}.

Note that one does not have to approximate $g_f$ as the sum of only power law segments $\Psi_s(m)$; other representations are possible too, as will be shown next.


\section{Example calculation: piecewise approximation of the fragment spectrum}
\label{sect:piecewise_example}


\begin{figure}
\centering
\includegraphics[width=0.5\textwidth]{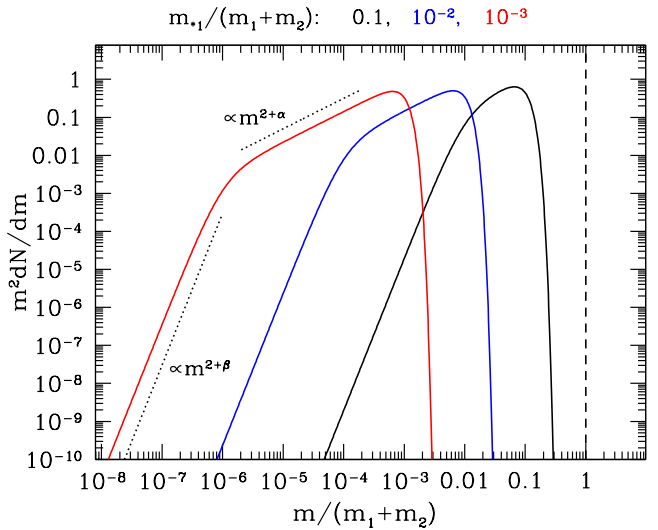}
\caption{
Fragment mass spectrum given by equation (\ref{eq:spectrum}), shown for three different values of $m_{*1}=0.1,10^{-2},10^{-3}$ (with corresponding $m_{*1}=10^{-2},10^{-4},10^{-6}$). We used $\alpha=-1.2$ and $\beta=2$ in this illustration. The color scheme is illustrated at the top. Note the non-self-similar shape of the spectrum revealed by its evolution as $m_{*1}$ changes.
\label{fig:spectrum}}
\end{figure}

To demonstrate the accuracy and speedup associated with using our $\mathcal{O}\left(N^2\right)$ algorithm, we now provide an example of applying it to treat collisional evolution with a non-self-similar fragment size distribution. We consider a fragment mass spectrum
\ba
g_f(m|m_1,m_2) &=& A\exp\left[-\left(\frac{m}{m_{*1}}\right)^3\right]\left(\frac{m}{m_{*1}}\right)^\alpha
\nonumber\\
&\times & \left[1+\left(\frac{m_{*2}}{m}\right)^2\right]^{(\alpha-\beta)/2},
\label{eq:spectrum}
\ea
with normalization $A$, two mass scales, $m_{*1}$ and $m_{*2}$, and two power law slopes $\alpha$ and $\beta$. This mass spectrum is exponentially truncated above $m_{*1}$. It behaves as a power law $m^{-\alpha}$ for $m_{*2}\lesssim m\lesssim m_{*1}$, however, the power law slope smoothly changes to $\beta$ for very small fragments, $m\lesssim m_{*2}$. 

Very importantly, $m_{*1}/m_{*2}$ is not a constant but changes as the collision characteristics (e.g. masses $m_1$ and $m_2$) vary. This makes $g_f(m|m_1,m_2)$ given by equation (\ref{eq:spectrum}) non-self-similar, which is illustrated in Figure \ref{fig:spectrum}. There we show how the shape of $g_f(m|m_1,m_2)$ (multiplied by $m^2$) evolves as $m_{*1}$ and $m_{*2}$ vary, implying lack of self-similarity. Each of the curves is normalized such that the total mass in fragments is always $m_1+m_2-\epsilon m_{\rm rm}$ (although in this figure we set $\epsilon=0$ for simplicity).

\begin{figure}
\centering
\includegraphics[width=0.5\textwidth]{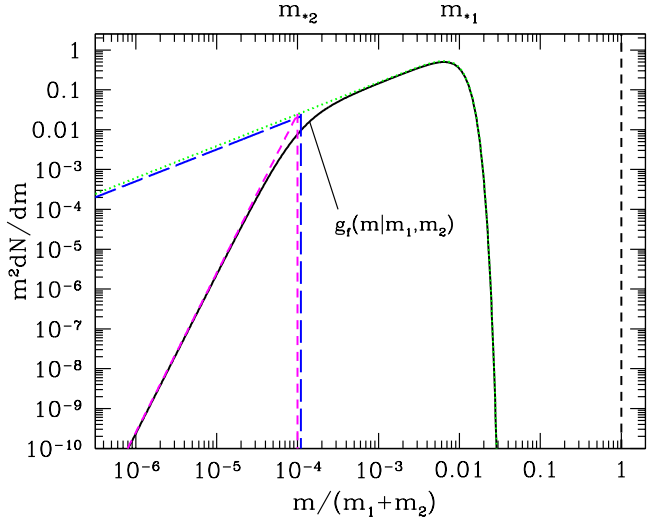}
\caption{
Decomposition of the fragment mass spectrum  (\ref{eq:spectrum}) into three different self-similar components, as shown in equation  (\ref{eq:decomposition}). See text for details.
\label{fig:spec_decomp}}
\end{figure}

\begin{figure*}
\centering
\includegraphics[width=\textwidth]{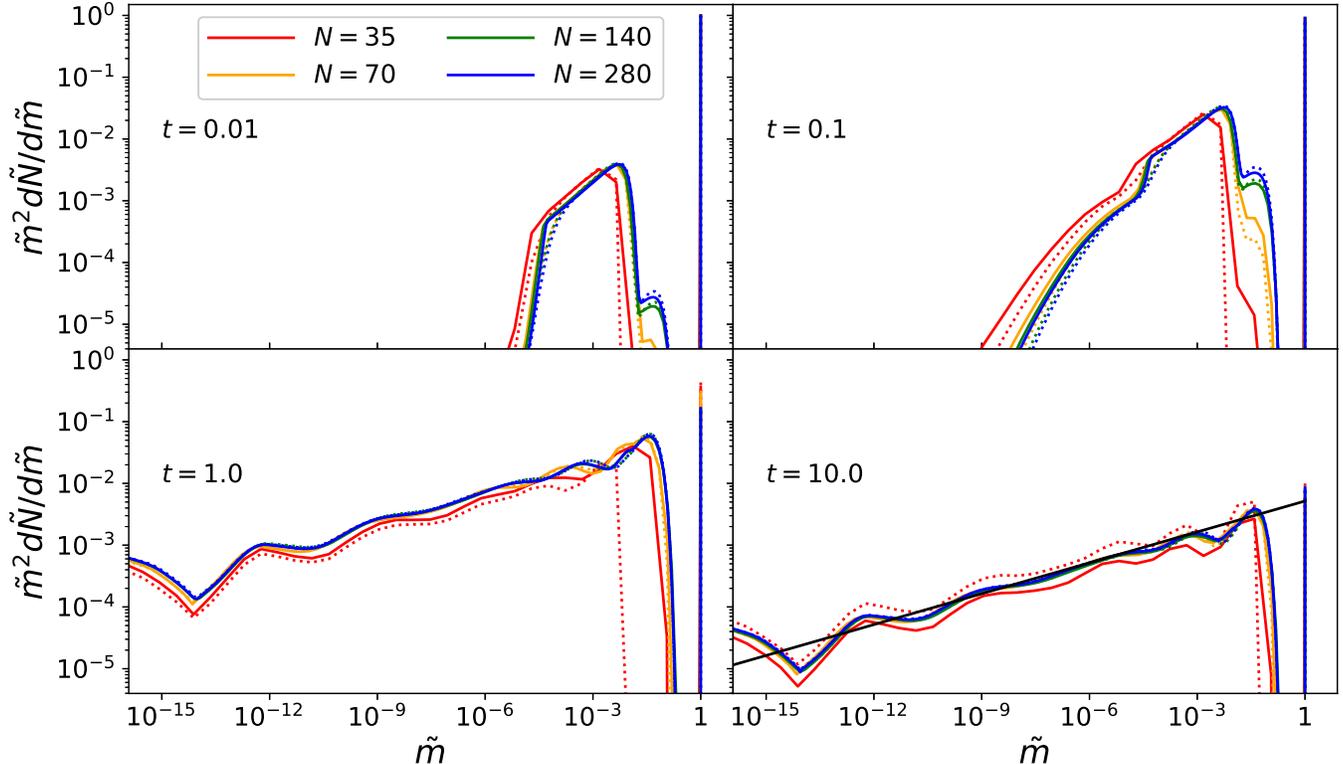}
\caption{
Mass spectrum at 4 different times indicated on the panels, expressed in units of collisional time scale for the initial mass bin.  Different color curves correspond to different numbers of bins in the simulation, labeled in panel (a).  Solid lines correspond to the curves from the $\mathcal{O}\left(N^2\right)$ algorithm, and dotted lines to those from the $\mathcal{O}\left(N^3\right)$ algorithm. We use $\tilde m$ and $\tilde N$ for $m$ and $N$ normalized by the initial mass ($m=1$) and number of objects in the system.  In the bottom right panel, we show in black the line of slope 1/6 (consistent with \citealt{Dohnanyi69}), which provides a good overall fit to the fragment distribution. The superimposed wavy structure is discussed in the text.
\label{fig:comparison}}
\end{figure*}

\begin{figure}
\centering
\includegraphics[width=0.5\textwidth]{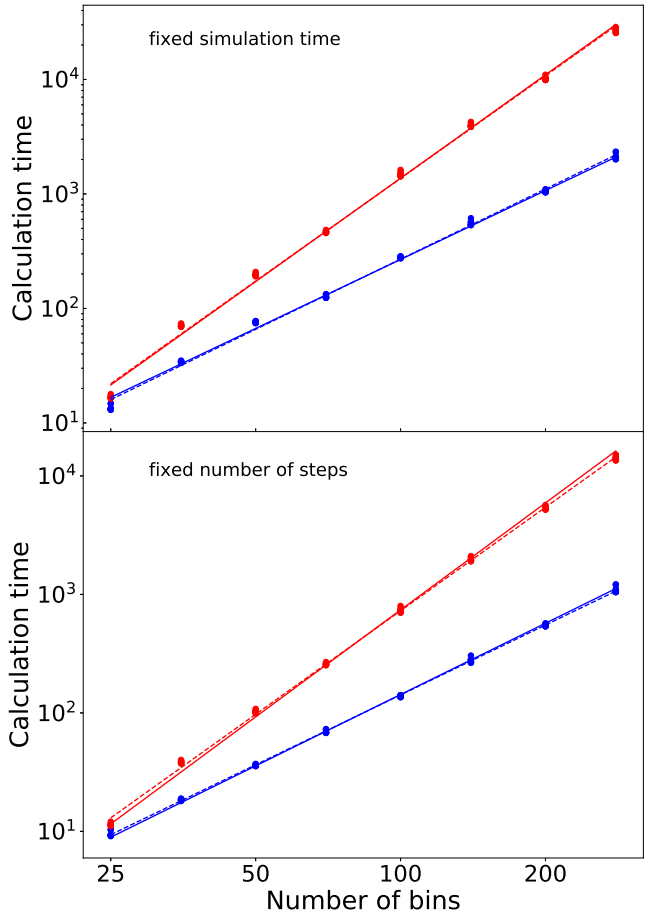}
\caption{
Run time as a function of the number of bins, expressed in arbitrary units.  The data points in the top panel show the amount of time needed to reach a simulation time of $t=1$ in units of initial collision times.  The data points in the bottom panel show the time needed to run for 400 steps.  The blue points and curves are for the $\mathcal{O}(N^2)$ algorithm, and the red ones for the $\mathcal{O}(N^3)$ algorithm.  The solid lines are the best fit power law to all the points, assuming the slopes to be exactly 2 and 3.  The dotted lines are the best fit power laws determined without fixing the slope.  See text for more details.
\label{fig:timings}}
\end{figure}

We can approximate the mass spectrum (\ref{eq:spectrum}) as a superposition of three self-similar components as follows:
\ba
&& g_f(m|m_1,m_2)\approx  A\exp\left[-\left(\frac{m}{m_{*1}}\right)^3\right]\left(\frac{m}{m_{*1}}\right)^\alpha\nonumber\\
 &&  -  A_- \varphi^{\rm pl}\left(\frac{m}{m_{*2}}\Big|\alpha\right)
 +A_+ \varphi^{\rm pl}\left(\frac{m}{m_{*2}}\Big|\beta\right),
 \label{eq:decomposition}
\ea
each of them featuring only one mass scale. This decomposition is illustrated in Figure \ref{fig:spec_decomp}. The first term (green dotted curve) is an exponentially truncated power law extending all the way down to very small fragment sizes; it is designed to fit the original spectrum (\ref{eq:spectrum}) for $m\gtrsim m_{*2}$. The second term (blue dashed line) is a power law with the same slope $\alpha$ as in the first term, sharply truncated above $m_{*2}$. Its amplitude $A_-$ is chosen so that it fully offsets the first term below $m_{*2}$ (note that it enters with the negative sign). Finally, the last component (dashed magenta line) is another power law sharply truncated at $m_{*2}$ with the slope $\beta$ and amplitude $A_+$ chosen such that this term matches the behavior of the spectrum (\ref{eq:spectrum}) for $m\lesssim m_{*2}$.

We now carry out two fragmentation calculations. One uses fragment mass spectrum (\ref{eq:spectrum}) without approximations; because of its non-self-similar shape this calculation employs the standard $\mathcal{O}(N^3)$ fragmentation algorithm. The second calculation uses an approximation (\ref{eq:decomposition}), allowing us to use our $\mathcal{O}(N^2)$ algorithm as described in \S \ref{sect:piecewise}. Both of them use explicit time stepping (see Section \ref{sect:implicit} for comparison with implicit calculations). In both cases, we evolve the system using Euler's method.  The time-step is chosen so that the number of particles in one bin will not change by more than 10 \% in any one time-step (with an allowance for bins with a small number of particles in them). We then compare the outcomes of the two calculations, as well as the numerical costs involved.

In both cases we assume that $m_{*1}$ is given by 
\ba
m_{*1} = \frac{m_1}{10^2} \left(\frac{m_1}{m_2}\right)^{1/2},
\label{eq:m1}
\ea
where $m_1 \geq m_2$. We also choose 
\ba  
m_{*2}=\frac{m_{*1}^2}{m_1+m_2},
\label{eq:m2}
\ea
so that when $m_{*1}/(m_1+m_2)$ goes down (e.g. for more energetic collisions), there is a larger range in $\ln m$, for which $g_f(m|m_1,m_2)\propto m^\alpha$ (i.e. more small fragments get formed). We use $\alpha=-1.5$ and $\beta=2.5$ in this calculation. At time $t=0$ all mass in the system is in objects with the same mass $m=1$ (monodisperse initial condition) occupying a single mass bin. The mass interval that we cover extends from $m=10^{-16}$ to $m=1$; fragments falling below the lower mass end get removed from the system. We assume for simplicity that collisions lead to fragmentation only if $m_2/m_1 \geq 10^{-2}$. Also, we assume that no largest remnant remains, i.e. only the continuous spectrum of small fragments results in a fragmentation event. The collision rate is proportional to the geometric cross-section of the two colliding bodies, assuming all objects to be spheres of the same density. This setup is similar to that in \citet{Dohnanyi69} and \citet{Tanaka1996}, except for the non-self-similar shape of the fragment spectrum.

Results of the two calculations are shown in Figure \ref{fig:comparison}.  There we plot the mass distributions $dN/dm$ (multiplied by $m^2$ and normalized to the initial mass and particle number) at different times during the calculation, for the two different algorithms run with different numbers of mass bins.  Time, labelled on the panels, is in terms of the initial timescale for collisions between bodies in the initial mass bin. The height of the bin at $m=1$ shows the current number of particles in the initial mass bin normalized to the initial number of objects in the system. 

We first discuss the general features of the collisional evolution in this calculation. Early on, at $t=0.01$, the mass spectrum closely mirrors that of the assumed fragment size distribution (\ref{eq:spectrum}). This is to be expected, since at that time the number of objects with $m<1$ is small enough (total mass in this part of the spectrum is $\lesssim 1\%$) for their mutual collisions to be rare. On the other hand, collisions between these fragment and the numerous large $m=1$ objects do occur, which explains a bump\footnote{According to equation (\ref{eq:m1}), collisions with smaller fragments result in larger $m_{*1}$. Because of our assumption of no fragmentation when $m_2/m_1<10^{-2}$, the largest possible $m_{*1}$ is $0.1m_1$. This explains the gap between the initial mass bin at $m=1$ and the continuous spectrum of fragments, which persists at all times.} appearing above $m=10^{-2}$. This bump becomes more pronounced at $t=0.1$ and fully morphs with the continuous mass spectrum by $t=1$. 

By $t=0.1$ the shape of the distribution of fragments starts to evolve away from the single-collision spectrum (\ref{eq:spectrum}) at small $m$. And by $t=1$ the fragment mass distribution attains a steady-state form, which can be viewed as a power law with superimposed wavy structure. The slope of this power law is close to 1/6 (shown in black in the bottom right panel), in agreement with the results of \citet{Dohnanyi69} and \citet{Tanaka1996}. The wiggles on top of this power law are caused by the boundary condition at the low mass end, see \citet{Campo1994} for a discussion of this effect. Beyond $t=1$ only the normalization of the size distribution changes, steadily decaying in time because of mass lost to particles smaller than our smallest mass bin, while its overall shape stays the same. The height of the $m=1$ bin goes down too. It drops by two orders of magnitude by $t=10$, signaling substantial erosion of the initial population of objects.


\subsection{Comparison of the $\mathcal{O}\left(N^2\right)$ and $\mathcal{O}\left(N^3\right)$ algorithms}
\label{sect:compare}


We now compare the performance of the $\mathcal{O}\left(N^2\right)$ algorithm and the full $\mathcal{O}\left(N^3\right)$ calculation. We first note that for small numbers of bins ($N=35$), there are substantial differences between the results of the two calculations at all times.  However, as the number of bins increases, the results converge, with two algorithms agreeing with each other quite well already for $N=70$. Minor differences remain, especially at early times (when there is still a strong sensitivity to the shape of the input fragment spectrum), as even in the limit of an infinite number of bins, the fragment mass distributions are slightly different near $m_{*2}$ (see Figure \ref{fig:spec_decomp}). This can be seen, for example, near $m=10^{-5}-10^{-4}$ (right around $m_{*2}$ for collisions of two $m=1$ objects) at $t=0.01$ for $N=280$. Nevertheless, at late times these differences get largely wiped out. Thus, we can conclude that already with 5-10 mass bins per decade our $\mathcal{O}\left(N^2\right)$ algorithm is able to reproduce the fine details of the collisional evolution even for non-self-similar fragment size spectrum. 

Turning now to the computational cost of each algorithm, in Figure \ref{fig:timings} we show the wall clock time to run them as a function of the number of bins $N$.  We use an adaptive time-step, so the number of time-steps required to evolve the distribution for a given simulation time is slightly different between the two algorithms, and for different numbers of bins.  For that reason, we plot both the amount of time required to reach a fixed simulation time ($t=1$, upper panel), and the time required to execute 400 steps of the run (lower panel). 

There are slight variations in the run time even for the exact same parameters, presumably caused by the evolving state of the computing hardware. For this reason, for each algorithm and number of bins, we run the calculation 10 times, hence the multiple points shown in Figure \ref{fig:timings} for each number of bins.  We then calculate the best fit lines through the points for each algorithm, assuming $T\propto N^2$ and $N^3$ scalings, correspondingly.  These are the solid lines in the figure.  We also calculate the best fit lines without fixing their slopes, which are shown as the dotted lines.  These slopes turn out to be 2.04 and 2.98 in the top panel, and 1.96 and 2.90 in the bottom panel, in good agreement with the theoretical expectations.


\section{Implicit Versus Explicit Time Stepping}
\label{sect:implicit}


A number of studies have used implicit time integration methods to evolve the coagulation-fragmentation equations \citep{Brauer2008, Birnstiel2010, Garaud13, Booth2018}. The advantage of these methods has traditionally been that they allow much longer time steps to be used in the integration, leading to a faster time to solution despite the increased complexity of the method. However, these studies did not make use of the $\mathcal{O}(N^2)$ fragmentation algorithm, which can only be used with explicit time stepping.

\begin{figure}
\centering
\includegraphics[width=0.5\textwidth]{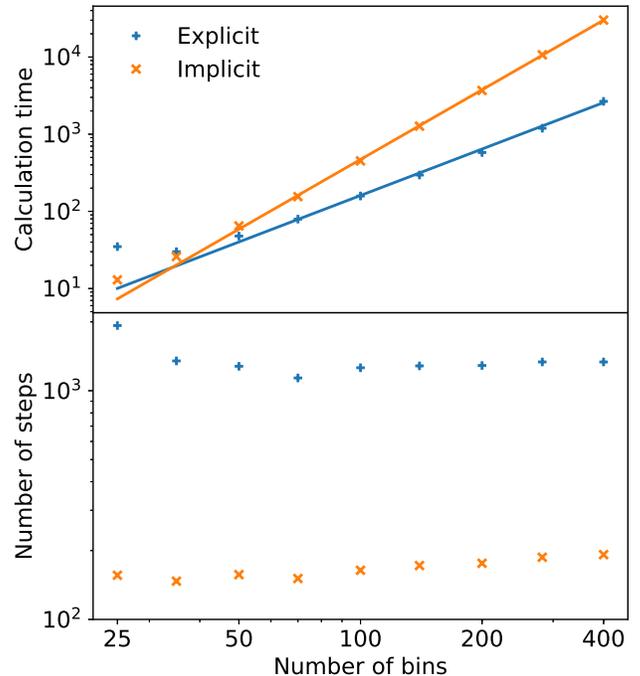}
\caption{
Comparison of the time required (top), shown in arbitrary units, and number of steps taken (bottom) to evolve the system for $10$ initial collisional time scales for the 3rd-order implicit (orange) and explicit (blue) methods. The lines show best fits assuming slopes of 2 and 3.}
\label{fig:implicit}
\end{figure}

The difference between explicit and implicit methods can be summarized as follows. Master equation (\ref{eq:master_m}) written in a discretized form in mass space has a form of a system of equations 
\begin{equation}
    \frac{\partial n_i}{\partial t} = f_i(t, \vec{n}),
\end{equation} 
where, as before, $\vec{n}=\{n_i\}$ and vector $\vec{f}=\{f_i\}$ stands for the expressions in the right hand side of equation (\ref{eq:master_m}) written out for each $i$. When evolving this system, explicit updates by $\delta t$ in time take the form 
\begin{equation}
    n_i(t + \delta t) = n_i(t) + f_i(t, \vec{n}(t)) \delta t,
\end{equation}
whereas implicit updates reduce to solving the system of equations
\begin{equation}
    n_i(t + \delta t) = n_i(t) + f_i(t+\delta t, \vec{n}(t+\delta t)) \delta t.
\end{equation}
for $n_i(t + \delta t)$. Writing $F(n_i^*)= n_i^* - n_i(t) - f_i(t+\delta t, \vec{n}^*)\delta t$, the solution $F(n_i(t+\delta t)) = 0$ can be found via Newton-Raphson iteration, in which the estimate $n_i^*$ is updated via
\begin{equation}
   n_i^* \rightarrow  n_i^* - \left[\hat I - \frac{\partial \vec{f} }{\partial \vec{n}} \delta t \right]^{-1} F(n_i^*), 
   \label{eq:implicit_update}
\end{equation}
where $\hat I$ is the identity matrix. 

The appearance of the Jacobian, ${\partial \vec{f} }/{\partial \vec{n}}$, and matrix inversion in the above equation prevents implicit methods from benefiting from the $\mathcal{O}(N^2)$ computation of the fragmentation rate because both the Jacobian computation and the matrix inversion in equation~(\ref{eq:implicit_update}) require $\mathcal{O}(N^3)$ operations to compute. The $\mathcal{O}(N^3)$ complexity of the Jacobian calculation can be understood from equation~(\ref{eq:Af}) as ${\partial \vec{\mathcal{A}}}/{\partial \vec{n}}$ is an $N\times N$ matrix.

To compare the efficiency of implicit $\mathcal{O}(N^3)$ and explicit $\mathcal{O}(N^2)$ schemes, we use a simplified version of the problem presented in section~\ref{sect:compare}. We take the fragment spectrum to have a self-similar form
\ba
g_f(m|m_1,m_2) =  A\exp\left[-\left(\frac{m}{m_{*1}}\right)^3\right]\left(\frac{m}{m_{*1}}\right)^\alpha,
\ea
with $m_{*1}$ defined by equation (\ref{eq:m1}) as before. In this case, to integrate these equations we choose two 3rd-order methods, the explicit 3rd-order Runge-Kutta method of \citet{Gottlieb1998} and the implicit 3rd-order Rosenbrock method of \citet{Rang2005} used by \citet{Booth2018}. Both of these methods provide an embedded error estimates, which are used to adapt the time step to ensure that the relative error is below 1 per cent. For the implicit scheme we use the traditional $\mathcal{O}(N^3)$ fragmentation algorithm while the $\mathcal{O}(N^2)$ scheme is used with the explicit time integration scheme.

The time taken and number of steps required by the schemes to integrate the fragmentation equations to $t=10$ in units of initial collisional timescales are shown in Fig.~\ref{fig:implicit}. While the implicit method requires a factor 7 -- 9 fewer steps than the explicit scheme (bottom panel), the extra cost of the $\mathcal{O}(N^3)$ algorithm outweighs this for problems with more than about 50 cells (top panel). Tests on problems including both coagulation and fragmentation lead to similar conclusions.


\section{Discussion}  
\label{sect:sum}


The main result of this work is the $\mathcal{O}\left(N^2\right)$ algorithm for numerical treatment of fragmentation in collisional systems, described in \S \ref{sect:N2frag}. The main condition necessary for this approach to work is that the continuous size distribution of fragments resulting in an individual collision $g_f(m|m_1,m_2)$ depends on a single mass scale $m_*$. This algorithm is insensitive to the details of the actual dependence of the different characteristics of the fragment mass spectrum --- $m_{\rm rm}$, $m_*$, and $A$, see equations (\ref{eq:coll_outcome})-(\ref{eq:ss_model}) --- on the energy and masses of objects involved in a collision.  

Self-similarity of $g_f(m|m_1,m_2)$ is not a highly demanding requirement since the majority of numerical studies of fragmentation in astrophysical systems use such self-similar size distributions of fragments anyway, typically in the form of a truncated power law (\ref{eq:coll_outcome_pl}), see \citet{Greenberg1978}, \citet{Kenyon1999}, \citet{Lohne2008}, \citet{Brauer2008}, \citet{Birnstiel2010}. On the other hand, implementation of this algorithm allows substantial gains in computational efficiency, significantly reducing the time consumed by fragmentation simulations with large number of mass bins, $N\gtrsim 10^2$, see \S  \ref{sect:compare}. 

Moreover, as we showed in \S \ref{sect:generalize}-\ref{sect:piecewise_example}, the $\mathcal{O}\left(N^2\right)$ algorithm can be applied even when the fragment size spectrum is not a simple self-similar function with a single mass scale; one just need to approximate the non-self-similar fragment size distribution using several self-similar components. A practical example shown in \S \ref{sect:piecewise_example} demonstrates that the differences between a calculation carried with this algorithm and the direct `exact' $\mathcal{O}\left(N^3\right)$ calculation, which takes much longer, are very minor (at the level of several per cent or less) in systems that have evolved for longer than their characteristic collisional timescale. At early times the level of agreement is dictated mainly by the accuracy with which the original complicated spectrum of fragments is approximated by the self-similar components.

There is a reason why approximating even rather complicated non-self-similar fragment size distributions (e.g. measured in some experiments, \citealt{Fujiwara1977}) with a simple self-similar shape works in practice. The characteristics of the steady-state collisional cascades are known to be rather insensitive to the input fragment size spectrum $g_f(m|m_1,m_2)$. For example, \citet{Tanaka1996} has shown that as long as $g_f(m|m_1,m_2)$ is self-similar with $m_*\propto m_1$, the slope of the steady-state cascade should be sensitive only to the scaling of the collision rate with the masses of objects involved in a collision, but not to the actual form of $g_f(m|m_1,m_2)$. Similarly, \citet{OBrien2003} have shown that the slope of the collisional cascade depends on the mass scaling of the energy necessary to disrupt an object, but not on the power law of the fragment size spectrum (as long as $m_*\propto m_1$). By abandoning the assumption $m_*\propto m_1$, \citet{Belyaev2011} were able to demonstrate the sensitivity of the steady-state cascade to the shape of the fragment size spectrum; however, the variation was found to be very weak (logarithmic). This is one of the reasons why on long time intervals, after several collisional timescales have passed and the system settled into a steady-state cascade, our $\mathcal{O}\left(N^2\right)$ algorithm performs just as well as the exact calculation, see Figure \ref{fig:comparison}c,d.

Fragmentation algorithms documented in the literature (e.g. \citealt{Greenberg1978, Kenyon1999,Lohne2008,Brauer2008,Windmark2012,Garaud13}, etc.) redistribute the debris produced in collisions in the direct manner as described in \S \ref{sect:intro} and are thus $\mathcal{O}\left(N^3\right)$. To the best of our knowledge, \citet{Booth2018} is the only other study mentioning the possibility of constructing $\mathcal{O}\left(N^2\right)$ algorithm for self-similar fragment size distributions, however, without providing details. Our present study is intended partly to fill this gap. 

A number of studies invoke implicit time integration \citep{Brauer2008, Birnstiel2010, Garaud13}, which has an $\mathcal{O}\left(N^3\right)$ complexity due to the Jacobian calculation. The benefit of these schemes has traditionally been that the larger time steps they allow outweigh the additional cost in solving the linear system, which is only a factor $\sim 2$ when already using an $\mathcal{O}\left(N^3\right)$ fragmentation algorithm. Since our $\mathcal{O}\left(N^2\right)$ algorithm is about an order of magnitude faster than an implicit method per step already for $N=100$, this means that the time-step for implicit methods must be smaller by a similar factor to remain competitive. Although \citet{Brauer2008} did achieve a reduction in the number of time steps by a factor $\sim 100$ by using implicit methods for a problem including both grain growth and radial drift, this was primarily because radial drift limits the time-step in the explicit code to smaller values than those required by coagulation/fragmentation calculation. Without computing radial drift implicitly, the explicit $\mathcal{O}\left(N^2\right)$ approach is substantially faster, as demonstrated in Fig. \ref{fig:implicit}. The simplicity of our $\mathcal{O}\left(N^2\right)$ algorithm also makes it easier to implement and parallelize, as well as using less memory than fully implicit methods (as the entire Jacobian need not be stored). This makes our algorithm more attractive for complex problems.

Recently, simulations of dust dynamics in protoplanetary disks started including evolution of the dust size distribution due to coagulation/fragmentation {\it spatially resolved} in multiple dimensions \citep{Li2019,Draz2019}. As this is done using the existing $\mathcal{O}\left(N^3\right)$ framework of \citet{Birnstiel2010}, there is an associated computational overhead that scales steeply with the number of mass bins used to characterize dust size distribution. Use of our $\mathcal{O}\left(N^2\right)$ fragmentation algorithm would substantially reduce the computational cost of such calculations, making this tool an attractive option for future (multi-dimensional) studies of the dust evolution in disks around young stars. 

\acknowledgements
Financial support for this study has been provided by NASA via grant 15-XRP15-2-0139. RRR and RB acknowledge support from the STFC consolidated grants ST/P000673/1 and ST/S000623/1, respectively.


\bibliographystyle{apj}
\bibliography{references}





\end{document}